\documentclass[12pt]{article}
\usepackage{amssymb,amsmath,epsfig}
\usepackage{cite}
 \allowdisplaybreaks

\begin{document}
\title{\bf Influence of Modification of Gravity on the Complexity Factor of Static Spherical Structures}
\author{Z. Yousaf$^1$ \thanks{zeeshan.math@pu.edu.pk}, Maxim Yu. Khlopov$^{2,3,4}$ \thanks{khlopov@apc.in2p3.fr},\
M. Z. Bhatti$^1$ \thanks{mzaeem.math@pu.edu.pk} and T. Naseer$^1$ \thanks{tayyabnaseer48@yahoo.com}\\
$^1$Department of Mathematics, University of the Punjab,\\
Quaid-i-Azam Campus, Lahore-54590, Pakistan.\\
$^2$Institute of Physics, Southern Federal University,\\ 194
Stachki, Rostov-on-Donu, Russia.\\
$^3$ Université de Paris,
CNRS, Astroparticule et Cosmologie,\\ F-75013 Paris, France.\\
$^4$ National Research Nuclear University MEPhI \\ (Moscow
Engineering Physics Institute), 115409 Moscow, Russia}

\date{}

\maketitle
\begin{abstract}
The aim of this paper is to generalize the definition of complexity
for the static self-gravitating structure in $f(R,T,Q)$
gravitational theory, where $R$ is the Ricci scalar, $T$ is the
trace part of energy momentum tensor and $Q\equiv
R_{\alpha\beta}T^{\alpha\beta}$. In this context, we have considered
locally anisotropic spherical matter distribution and calculated
field equations and conservation laws. After the orthogonal
splitting of the Riemann curvature tensor, we found the
corresponding complexity factor with the help of structure scalars.
It is seen that the system may have zero complexity factor if the
effects of energy density inhomogeneity and pressure anisotropy
cancel the effects of each other. All of our results reduce to
general relativity on assuming $f(R,T,Q)=R$ condition.
\end{abstract}

\section{Introduction}

The general relativity (GR) was proposed by Albert Einstein in 1915
in which he related matter and space-time through Einstein field
equations. General Relativity could be recognized as the basis for
gravitational physics and cosmology. It can used to express the
history and expansion of our universe, the black hole phenomena and
light that comes from distant galaxies. To get some feasible results
about our universe on different scales, the $f(R)$ theory of gravity
was introduced by replacing the Ricci scalar $R$ with its generic
function in an action function. Nojiri and Odintsov
\cite{nojiri2007introduction} considered $f(R)$ gravity and
described different phases of our universe like cosmological
structure and phantom era. Capozziello \emph{et al.}
\cite{PhysRevD.83.064004} considered the Lan\'{e}-Emden equation
with $f(R)$ corrections in order to study the hydrostatic
equilibrium of stellar objects. They also found a mathematical
connection of pressure with density and compared their results with
that obtained in GR. Bamba \emph{et al.} \cite{bamba2012dark}
explored the properties of different cosmologies with dark energy.
They also investigated the $\Lambda$CDM-like universe by considering
different cosmological models. Various researchers
\cite{cembranos2012gravitational,sharif2014instability,dymnikova2015regular,
bhatti2017gravitational,abbas2018complexity,doi:10.1142/S0217732319503334,bhatti2020stability,yousaf2020construction,yousaf2020hamna}
investigated the effects of curvature terms on the formation and
evolution of self-gravitating structures. After obtaining some exact
solutions of $f(R)$ field equations, they also highlighted some
applications of their results. Harko \emph{et al.} \cite{harko2011f}
introduced the $f(R,T)$ theory of gravity which can be considered as
the extension of $f(R)$ theory, where $T$ denotes the trace part of
energy-momentum tensor. They presented the modified field equations
in $f(R,T)$ gravity and analyzed the motion of test particles
through variational principle. Baffou \emph{et al.}
\cite{baffou2015cosmological} proposed a model to study the dynamics
and stability of this theory through de Sitter and power-law
solutions. They obtained few observationally compatible cosmological
solutions.

The $f(R,T,Q)$ theory is based on the non-minimal coupling (NMC)
between matter and geometry. In order to clarify the role of dark
matter and dark energy in any stellar object without restoring to
exotic matter, the formation of Einstein-Hilbert action is modified.
Initially, the geometric part in the action was modified by
replacing the Ricci scalar with its generic function. Later, $f(R)$
gravity was found not to meet the standard solar system constraints
\cite{chiba20031,olmo2007limit}. Consequently, Harko \emph{et al.}
\cite{harko2011f} introduced the $f(R,T)$ theory of gravity which is
the extension of $f(R)$ theory, where $T$ denotes the trace part of
energy-momentum tensor. It is important to mention that the $f(R,T)$
gravity could not be able to encompass the NMC effects in the
gravitational equations for $T=0$, while an additional term
$T^{\alpha\beta}R_{\alpha\beta}$ could provide these effects in this
context. Haghani \emph{et al.} \cite{haghani2013further} studied the
role of strong NMC between geometry and fluid distribution by
studying $f(R,T,Q)$ theory (where $Q\equiv
T^{\alpha\beta}R_{\alpha\beta}$), which could be regraded as the
generalization of $f(R,T)$ gravity. The fulfilment of the solar
system tests and stability criteria are the fundamental requirements
of any gravitational theory. Usually, the energy-momentum tensor is
not conserved in $f(R,T,Q)$ theory. Haghani \emph{et al.}
\cite{haghani2013further} found the stability conditions for this
theory by interpreting the Dolgov-Kawasaki instability. In this kind
of theories, the conservation of the stress energy tensor is also
possible and in this case, throughout the high density era of our
cosmic evolution, the existence of a de Sitter phase was to be
found. They also utilized the Lagrange multiplier method to find the
gravitational equations with conserved energy-momentum tensor.

Due to the non-conserved nature of this gravity and matter geometry
interaction, an additional force is always present, even in our
considered case $L_{m}=-\mu$, where $\mu$ is the energy density of
the fluid, and thus the motion of particles does not follow geodesic
path. The existence of an additional force could be helpful to study
the galactic properties. The fluid-geometry coupling in $f(R,T,Q)$
theory may help us to analyze the reason of the late time
acceleration of our cosmos. The cosmological aspects and the
accelerating solutions of this theory has also been investigated by
Odintsov and S\'{a}ez-G\'{o}mez \cite{odintsov2013f}. The results of
Dolgov-Kawasaki instability calculated in \cite{haghani2013further}
and \cite{odintsov2013f} are found to be same. Elizalde and Vacaru
\cite{elizalde2015effective} considered some cosmological models and
constructed off-diagonal analytical solutions in $f(R,T,Q)$ gravity.
They also studied the FLRW cosmological model as well as
$\Lambda$CDM universe, and the nonholonomic cosmological solutions
are also described.

Haghani \emph{et al.} \cite{haghani2014matter} considered $f(R,T,Q)$
gravity and found some stability conditions regarding local
perturbations. They examined the cosmological consequences which
provide an exponential solution, and concluded that in the
gravitational dynamics, the matter itself may play a key role. They
also shown that de Sitter type solutions can also be conceded by
modified field equations. Gama \emph{et al.} \cite{gama2017godel}
dealt with the $f(R,Q)$ gravity in order to calculate G\"{o}del-type
solutions and compared there stability with the GR solutions. They
considered a particular model in this gravity to obtain causal
solutions and found some conditions of their existence in the light
of matter sources. They noticed that the rotating universe can be
described by the G\"{o}del-type metric, but its expansion was not
taken into account. This theory may help us to explore the new
aspects of the earliest stages of our cosmic evolution in near
future.

Odintsov and S\'{a}ez-G\'{o}mez \cite{odintsov2013f} determined
numerical as well as analytical solutions of $f(R,T,Q)$ theory and
checked the correspondence of their solutions with $\Lambda$CDM
model. They also explained solutions for de Sitter universe and
problems containing fluid instability. Ayuso \emph{et al.}
\cite{ayuso2015consistency} studied the consistency and the
stability of $f(R,T,Q)$ theory with an appropriate scalar/vector
field. They produced higher order equations of motion describing the
matter fields through the conformal and non-minimal couplings.
Baffou \emph{et al.} \cite{baffou2016exploring} discussed the
stability through the de Sitter and power-law solution to explain
the early evolution of our universe. They found numerical solution
of some special models of $f(R,T,Q)$ theory to discuss their
stability. In the gravitational collapse of spherical structure,
Bhatti \emph{et al.} and his collaborators
\cite{yousaf2016stability,Yousaf2017,yousaf2017stability,bhatti2019dissipative}
analyzed the role of the physical variables on the dynamical
evolution of self-gravitating systems and found the relation of
structural variables with that of Weyl tensor in $f(R,T,Q)$ theory.

A complexity is the combination of various components which could be
a source to trigger complications in any stable and static
self-gravitating system. The definition of complexity has been
examined in different fields of science. There are many definitions
of complexity and one of them is introduced by L\'{o}pez-Ruiz
\emph{et al.}
\cite{lopez1995statistical,calbet2001tendency,catalan2002features}
that was based on the concepts of entropy and information. Entropy
gives us the disorderness of a system and information can be
familiarty about any system. There are many other components to
define the complexity of a system. The more appropriate way to
define complexity was suggested by L\'{o}pez-Ruiz \emph{et al.}
\cite{lopez1995statistical} which was based on the idea of
disequilibrium.

Another way to interpret the definition of complexity in physics
starts by dealing with isolated ideal gas and perfect crystal, as
these systems have zero complexity. The isolated ideal gas is a
system made-up of random moving molecules, so it is completely
scattered. It gives us maximum information because all molecules
participate equally. On the other hand, a perfect crystal is one
whose constituents are organized in a highly ordered form and it is
enough to study the small portion to describe its nature and hence
it gives less data information. These two models are extreme in
order and information. Hence, there would be maximum disequilibrium
in case of perfect crystal and zero in case of the isolated ideal
gas. As, they have zero complexity, so there is no complication in
the behavior of these systems. In astrophysics, the complexity
factor is significant in order to study the structure of the
self-gravitating systems. In this scenario, the components which
have been investigated usually are pressure, equilibrium, energy
density and luminosity. In the absence of pressure factor in the
energy-momentum tensor, energy density is not enough to express
complexity.

Lloyd and Pagels \cite{lloyd1988complexity} studied the observable
states of self-gravitating structures and found their complexity
factor that could be applicable to all physical structures. They
also found the complexity factor for computational systems as a
special case and discussed applications of some mathematical and
physical problems. Crutchfield and Young
\cite{crutchfield1989inferring} introduced the complexity of
nonlinear dynamical structures. They proposed a measure of
complexity of a system through entropy and dimension by a method
that recreate minimal equations of a structure. Herrera \emph{et
al.} \cite{herrera1998role} investigated the aspects of
inhomogeneous energy density and local pressure anisotropy on the
spherical collapse of matter distribution and explained the active
gravitational mass.

Herrera and Barreto \cite{herrera2013general} formulated polytropic
spherical structures with pressure anisotropy and discussed their
applications. They calculated the Tolman mass to interpret some
properties of relativistic structures. Herrera \emph{et al.}
\cite{herrera2004spherically,herrera2011role} proposed the
orthogonal splitting of the Riemann curvature tensor and found some
structure scalars. These scalars are then associated with the basic
properties of matter distribution. They declared all possible
solution of Einstein field equations in terms of these scalars
through some examples in static case \cite{herrera2009structure}.
Yousaf and his collaborators
\cite{Yousaf2019,yousaf2019role,Yousaf2018,yousaf2016electromagnetic,phys1}
as well as Sharif and Manzoor
\cite{sharif2015structure,sharif2017dark} extended these results for
various cosmic models in modified theories.

Thirukkanesh and Ragel \cite{thirukkanesh2012exact} studied
spherically symmetric static geometry and interpreted compact
structures of fluid distribution having pressure anisotropy. They
obtained some exact models by writing another form of field
equations using polytropic equation of state. Di Prisco \emph{et
al.} \cite{di1997cracking} dealt with two classes of homogeneous
matter distribution in which the fluctuations induced by local
pressure anisotropy may cause the cracking due to disequilibrium in
spherical compact objects.

The aim of this work is to present the complexity factor for the
locally anisotropic spherical matter configurations in $f(R,T,Q)$
gravity. We shall study the role of $f(R,T,Q)$ corrections in the
modeling of relativistic spherical structure through complexity
factor and structure scalars. The paper is outlined as under. In the
next section, we propose the physical variables and modified field
equations. Then we find one of the structure scalars known as the
complexity factor from the curvature tensor in Sec. 3. After this in
Sec. 4, we introduce the vanishing complexity factor condition and
provide two exact solutions of modified field equations. Finally, we
conclude all of our in Sec. 5.

\section{The Physical Variables and Other Equations}

We now model our system to be static spherically symmetric which is
coupled with anisotropic matter configurations. We study the
structure of such systems after considering $f(R,T,Q)$ equations of
motion. We shall also express our results with the help of Tolman
and Misner-Sharp formalisms. We describe various physical variables
involved in the description of a static self-gravitating fluids.
Further, we will evaluate few matching conditions on the three
dimensional boundary surface $\Sigma$.

\subsection{Modified Field Equations}

The action for $f(R,T,Q)$ theory is
\cite{odintsov2013f,ayuso2015consistency,baffou2016exploring}
\begin{equation}\label{a1}
S=\frac{1}{2}\int \sqrt{-g}\left[f(R,T,Q)+L_{m}\right]d^{4}x,
\end{equation}
where $L_{m}$ is the matter Lagrangian and defined as $L_{m}=-\mu$.
Here, $\mu$ denotes the energy density of the matter distribution,
$g$ describes the determinant of the metric tensor $g_{\alpha\beta}$

The field equations corresponding to above action can be written as
follows
\begin{equation}\label{a2}
G_{\mu\nu}=8\pi T_{\mu\nu}^{(eff)},
\end{equation}
where $G_{\mu\nu}$ stands for Einstein tensor and the term
$T_{\mu\nu}^{(eff)}$ could be regarded as the energy-momentum tensor
for $f(R,T,Q)$ theory, whose value can be given as follows
\begin{eqnarray}
\nonumber
T_{\mu\nu}^{(eff)}&=&\frac{1}{f_{R}-L_{m}f_{Q}}\left[\left(f_{T}+\frac{1}{2}Rf_{Q}+1\right)T_{\mu\nu}^{(m)}
+\left\{\frac{R}{2}(\frac{f}{R}-f_{R})-L_{m}f_{T}
\right.\right.\\\nonumber
&-&\left.\frac{1}{2}\nabla_{\alpha}\nabla_{\beta}(f_{Q}T^{\alpha\beta})\right\}g_{\mu\nu}-\frac{1}{2}\Box(f_{Q}T_{\mu\nu})
-(g_{\mu\nu}\Box-\nabla_{\mu}\nabla_{\nu})f_{R}\\\label{a3}
&-&2f_{Q}R_{\alpha(\mu}T_{\nu)}^{\alpha}+\nabla_{\alpha}\nabla_{(\mu}[T_{\nu)}^{\alpha}f_{Q}]+2(f_{Q}R^{\alpha\beta}
+\left.f_{T}g^{\alpha\beta})\frac{\partial^2 L_{m}}{\partial
g^{\mu\nu}\partial g^{\alpha\beta}}\right],
\end{eqnarray}
where $\nabla_\nu,$ describes the covariant derivation and
$\Box\equiv g^{\lambda\sigma}\nabla_\lambda\nabla_\sigma$. Moreover,
the subscripts $R,~T$ and $Q$ stand for partial differentiation with
respect to their arguments. The trace of stress-energy tensor in GR
provides a peculiar relationship between $R$ and $T$. However, in
this case, we found from Eq.\eqref{a3} as follows
\begin{align}\nonumber
&3\Box
f_R+\frac{1}{2}\Box(f_QT)-T(f_T+1)+\nabla_\pi\nabla_\rho(f_QT^{\pi\rho})+
R(f_R-\frac{T}{2}f_Q)\\\nonumber &+(Rf_Q+4f_T)\textit{L}_m
-2f+2R_{\pi\rho}T^{\pi\rho}f_Q -2
\frac{\partial^2\textit{L}_m}{\partial g^{\lambda\sigma}\partial
g^{\pi\rho}}\left(f_Tg^{\pi\rho}+f_QR^{\pi\rho}\right).
\end{align}
On assuming $Q=0$ in the above equation, one can observe
relativistic effects of $f(R,T)$ theory in the analysis, while the
consideration of vacuum case in this theory describes the dynamical
features of leads of $f(R)$ theory. The detailed analysis of their
derivation and physical implication in the study of our cosmic
structures are described in
\cite{odintsov2013f,ayuso2015consistency,baffou2016exploring}. In
Eq.\eqref{a3}, $T_{\mu\nu}^{(m)}$ is the usual energy-momentum
tensor which in our case can be written as
\begin{equation}\label{a6}
T_{\mu\nu}^{(m)}=\mu u_{\mu} u_{\nu}-Ph_{\mu\nu}+\Pi_{\mu\nu},
\end{equation}
where
\begin{equation}\label{a7}
\Pi_{\mu\nu}=\Pi\left(s_{\mu}
s_{\nu}+\frac{1}{3}h_{\mu\nu}\right);\quad
P=\frac{P_{r}+2P_{\bot}}{3},
\end{equation}
\begin{equation}\label{a8}
\Pi=P_{r}-P_{\bot};\quad h_{\mu\nu}=\delta_{\mu\nu}-u_\mu u_{\nu},
\end{equation}
and $s^{\mu}$ and $u^\mu$ are the four vectors, $\Pi$ is the
pressure anisotropy having radial pressure $P_{r}$ and tangential
pressure $P_{\bot}$ and $h_{\mu\nu}$ is the projection tensor.

We suppose that our geometry is characterized with a boundary
surface $\Sigma$ which has demarcated the interior and exterior
regions of spherical spacetimes. The geometry interior to $\Sigma$
can be given as follows
\begin{equation}\label{a10}
ds^2=-e^\lambda dr^2-r^2(d\theta^2+sin^2\theta d\phi^2)+e^\nu dt^2,
\end{equation}
where $\nu=\nu(r)$ and $\lambda=\lambda(r)$.

The four vectors corresponding to above system can be defined as
\begin{equation}\label{a4}
u^\mu=(e^{\frac{-\nu}{2}},0,0,0),\quad
s^\mu=(0,e^{\frac{-\lambda}{2}},0,0),
\end{equation}
from which one can write the four-acceleration as
$a^{\alpha}=u^{\alpha}_{;\beta}u^{\beta}$. We found only one
non-zero component of the four acceleration which can be given as
under
\begin{equation}\label{a5}
a_{1}=-\frac{\nu'}{2}.
\end{equation}
The 4-vectors satisfy the relations $s^\mu u_{\mu}=0, s^\mu
s_{\mu}=-1$.

The field equations in $f(R,T,Q)$ theory for the spherical system
\eqref{a6} and \eqref{a10} are
\begin{align}\label{a11}
&-\left[e^{-\lambda}\left(\frac{1}{r^2}-\frac{\lambda'}{r}\right)\right]+\frac{1}{r^2}=\frac{8\pi}{(f_{R}-L_{m}f_{Q})}\mu^{(eff)},\\\label{a12}
&-\left[-e^{-\lambda}\left(\frac{1}{r^2}-\frac{\nu'}{r}\right)\right]+\frac{1}{r^2}
=\frac{8\pi}{(f_{R}-L_{m}f_{Q})}P_{r}^{(eff)},
\\\label{a13}
&\frac{1}{32\pi}e^{-\lambda}\left[2\nu''+\nu'^2-\nu'\lambda'+{\frac{2(\nu'-\lambda')}{r}}\right],
=\frac{1}{(f_{R}-L_{m}f_{Q})}P_{\bot}^{(eff)},
\end{align}
where $\mu^{(eff)}, P_{r}^{(eff)}$ and $P_{\bot}^{(eff)}$ describe
the contribution of $f(R,T,Q)$ corrections in the physical variables
of relativistic fluids. Their values are given in Appendix A. Here,
prime indicates the derivative with respect to radial coordinate.

It is worthy to stress that, unlike GR and $f(R)$ theory, the
divergence of effective energy momentum tensor in $f(R,T,Q)$ gravity
is non-zero, which gives rise to the breaking of all equivalence
principles. Thus, the present theory encompasses non-geodesic motion
of the particles due to emergence of extra force acting on the
moving particles in this gravitational field. Its value can be
casted as
\begin{align}\label{14}
\nabla^\lambda
T_{\lambda\sigma}&=\frac{2}{Rf_Q+2f_T+1}\left[\nabla_\sigma(\textit{L}_mf_T)
+\nabla_\sigma(f_QR^{\pi\lambda}T_{\pi\sigma})-\frac{1}{2}(f_Tg_{\pi\rho}+f_QR_{\pi\rho})\right.\\\nonumber
&\times\left.\nabla_\sigma
T^{\pi\rho}-G_{\lambda\sigma}\nabla^\lambda(f_Q\textit{L}_m)\right].
\end{align}
For our observed system, the hydrostatic equilibrium can be studied
from the conservation equation as
\begin{equation}\label{a14}
\left(\frac{P_{r}^{(eff)}}{H}\right)'=\frac{-\nu'}{2H}(\mu^{(eff)}+P^{(eff)}_{r})
+\frac{2(P^{(eff)}_{\bot}-P^{(eff)}_{r})}{rH}+Ze^{\lambda},
\end{equation}
where $H=f_{R}-L_{m}f_{Q}$ and $Z$ indicates extra curvatures terms
of this theory described in Appendix A. This equation could be
called as the generalized Tolman-Opphenheimer-Volkoff (TOV) equation
for anisotropic matter which may help to understand the subsequent
changes in the structure of the static spherical system.

From Eq.\eqref{a12}, the value of $\nu'$ can be found as
\begin{equation}\label{a15}
\nu'=2\frac{m+4\pi r^3P^{(eff)}_{r}/H}{r(r-2m)}.
\end{equation}
The substitution of Eq.\eqref{a15} in Eq.\eqref{a14}, we get
\begin{align}\nonumber
\left(\frac{P_{r}^{(eff)}}{H}\right)'&=&-\frac{(m+4\pi
r^3P^{(eff)}_{r}/H)}{Hr(r-2m)}(\mu^{(eff)}+P^{(eff)}_{r})
+\frac{2(P^{(eff)}_{\bot}-P^{(eff)}_{r})}{Hr}+Ze^{\lambda},
\end{align}
where $m$ can be expressed through metric coefficient of the
spherical system as
\begin{equation}\label{a17}
R^3_{232}=1-e^{-\lambda}=\frac{2m}{r},
\end{equation}
which can be expressed through field equation \eqref{a11} as
\begin{equation}\label{a18}
m=4\pi\int_{0}^{r} \tilde{r}^2\frac{\mu^{(eff)}}{H} d\tilde{r}.
\end{equation}

We now describe the geometric structure outside $\Sigma$ with the
help of the spacetime given below
\begin{equation}\label{a19}
ds^2=\left(1-\frac{2M}{r}\right)dt^2-\frac{dr^2}{\left(1-\frac{2M}{r}\right)}-r^2(d\theta^2+\sin^2\theta
d\phi^2),
\end{equation}
where $M$ is a gravitating mass of the corresponding object. For the
smooth matching of outer and inner manifold across the boundary, we
consider Darmois junction conditions and matching criterion provided
by Yousaf \emph{et al.} \cite{Yousaf2017} (after following Senovilla
\cite{senovilla2013junction}) for $f(R,T,Q)$ gravity. At boundary
surface $r=r_{\Sigma}=$, we found that
\begin{align}\label{a20}
e^{\nu_{\Sigma}}=1-\frac{2M}{r_{\Sigma}},\quad
e^{-\lambda_{\Sigma}}=1-\frac{2M}{r_{\Sigma}},\quad
[P_{r}]_{\Sigma}=-D_{0},
\end{align}
where $D_{0}$ is given in Appendix A. However, the boundary
conditions for $f(R,T,Q)$ gravity are found to be
\begin{align}\label{27n1}
f_{,RR}[\partial_y R|_-^+=0,\quad
f_{,RR}K^*_{\lambda\sigma}|_-^+=0,\quad f_{,QQ}[\partial_y
Q|_-^+=0,\quad K|_-^+=0.
\end{align}
along with
\begin{align}\label{27n2}
R|_-^+=0,\quad Q|_-^+=0,\quad \gamma_{\lambda\sigma}|_-^+=0,
\end{align}
where $K^*_{ab}$ is the trace-free and and $K$ is the trace
component of the extrinsic curvature. The detailed analysis can be
found in \cite{Yousaf2017}. The conditions \eqref{27n1} and
\eqref{27n2} hold only, if $f_{,RR}\neq0$ and $f_{,QQ}\neq0$. Thus,
the boundary conditions (\ref{a19}) comes after applying the Darmois
matching condition, while the conditions \eqref{27n1} and
\eqref{27n2} arises due to $f(R,T,Q)$ theory. The satisfaction of
such constraints over $\Sigma$ are necessary for the smooth joining
of manifolds even for matter thin shells in $f(R,T,Q)$ gravity.

\subsection{Curvature Tensors}

It could be useful to express one of the well-known curvature
tensor, i.e., the Riemann tensor through the Ricci tensor
$R_{\alpha\beta}$, tensor $C_{\alpha\beta\mu\rho}$ and the Ricci
scalar $R$ as
\begin{eqnarray}\nonumber
R^\rho_{\alpha\beta\mu}&=&C^\rho_{\alpha\beta\mu}+\frac{1}{2}R^\rho_{\beta}g_{\alpha\mu}-\frac{1}{2}R_{\alpha\beta}\delta^\rho_{\mu}
+\frac{1}{2}R_{\alpha\mu}\delta^\rho_{\beta}-\frac{1}{2}R^\rho_{\mu}g_{\alpha\beta}\\\label{a23}
&-&\frac{1}{6}R\left(\delta^\rho_{\beta}g_{\alpha\mu}-g_{\alpha\beta}\delta^\rho_{\mu}\right).
\end{eqnarray}
The magnetic part of Conformal tensor vanishes for spherically
symmetric distribution, while its electric part
($E_{\alpha\beta}=C_{\alpha\gamma\beta\delta}u^{\gamma}u^{\delta}$)
can be given as
\begin{equation}\label{a24}
C_{\mu\nu\kappa\lambda}=(g_{\mu\nu\alpha\beta}g_{\kappa\lambda\gamma\delta}-\eta_{\mu\nu\alpha\beta}\eta_{\kappa\lambda\gamma\delta})u^\alpha
u^\gamma E^{\beta\delta},
\end{equation}
where
$g_{\mu\nu\alpha\beta}=g_{\mu\alpha}g_{\nu\beta}-g_{\mu\beta}g_{\nu\alpha}$,
and $\eta_{\mu\nu\alpha\beta}$ is the Levi-Civita tensor. We can
rewrite $E_{\alpha\beta}$ as
\begin{equation}\label{a25}
E_{\alpha\beta}=E\left(s_{\alpha}s_{\beta}+\frac{1}{3}h_{\alpha\beta}\right),
\end{equation}
with
\begin{equation}\label{a26}
E=-\frac{e^{-\lambda}}{4}\left[\nu''+\frac{\nu'^2-\lambda'\nu'}{2}-\frac{\nu'-\lambda'}{r}+\frac{2(1-e^\lambda)}{r^2}\right],
\end{equation}
and satisfy the following conditions
\begin{equation}\label{a27}
E^\alpha_{\alpha}=0,\quad E_{\alpha\gamma}=E_{(\alpha\gamma)},\quad
E_{\alpha\gamma}u^\gamma=0.
\end{equation}

\subsection{The Mass Function}

Here, we shall adopt the formalism provided by Misner-Sharp and
Tolman to calculate few interesting equation that would assist us to
study the structural properties of a relativistic fluid. After this,
we will find some connection between mass function and Conformal
tensor. Utilizing Eqs.\eqref{a2}, \eqref{a17}, \eqref{a23} and
\eqref{a25}, one can write
\begin{equation}\label{a28}
m=\frac{4\pi}{3H}r^3(\mu^{(eff)}+P^{(eff)}_{\bot}-P^{(eff)}_{r})+\frac{r^3E}{3},
\end{equation}
which can be manipulated as
\begin{equation}\label{a29}
E=-\frac{4\pi}{r^3}\int_{0}^{r}
\tilde{r}^3\left(\frac{\mu^{(eff)}}{H}\right)'
d\tilde{r}+\frac{4\pi}{H} (P^{(eff)}_{r}-P^{(eff)}_{\bot}).
\end{equation}
The above equation gives us the relation among Conformal tensor and
spherical structural properties, like effective energy density
inhomogeneity and effective form of the anisotropic pressure. By
making use of Eq.\eqref{a29} in Eq.\eqref{a28}, we have
\begin{equation}\label{a30}
m(r)=\frac{4\pi}{3H}r^3\mu^{(eff)}-\frac{4\pi}{3}\int_{0}^{r}
\tilde{r}^3 \left(\frac{\mu^{(eff)}}{H}\right)' d\tilde{r},
\end{equation}
which provides a peculiar relationship of the mass function with
homogeneous energy density. One can study the effects of effective
modified terms on the subsequent changes brought by energy density
inhomogeneity in the spherical anisotropic self-gravitating system
with the help of the above formula.

Tolman \cite{tolman1930use} introduced another formula of energy for
a static spherical distribution of matter given by
\begin{equation}\label{a31}
m_{T}=4\pi \int_{0}^{r_{\Sigma}}
r^2e^{(\nu+\lambda)/2}(T_{0}^{0(eff)}-T_{1}^{1(eff)}-2T_{2}^{2(eff)})dr.
\end{equation}
Bhatti \emph{et al.} \cite{zaeem2019energy,bhatti2019tolman}
calculated the expressions for Tolman mass function for the case
spherically symmetric systems in $f(R)$ gravity with and without
electromagnetic field. This was introduced to estimate the total
energy of the structure and within the spherical configuration of
radius $r$. It can be described as
\begin{equation}\label{a32}
m_{T}=4\pi \int_{0}^{r}
r^2e^{(\nu+\lambda)/2}(T_{0}^{0(eff)}-T_{1}^{1(eff)}-2T_{2}^{2(eff)})dr.
\end{equation}
Using Eqs.\eqref{a11}-\eqref{a13} in above equation, one may obtain
\begin{equation}\label{a33}
m_{T}=e^{(\nu+\lambda)/2}[m(r)+4\pi r^3P^{(eff)}_{r}/H].
\end{equation}
Using the definition of mass function and the field equations, one
can write
\begin{equation}\label{34}
m_{T}=e^{(\nu-\lambda)/2}\nu'\frac{r^2}{2}.
\end{equation}
This equation gives us the physical importance of $m_{T}$ as the
effective inertial mass. In a static gravitational field (instantly
at rest), the gravitational acceleration $(a=-s^{\nu}a_{\nu})$ of a
test particle is followed by
\begin{equation}\label{a35}
a=\frac{e^{-\lambda/2}\nu'}{2}=\frac{e^{-\lambda/2}m_{T}}{r^2}.
\end{equation}
The more suitable representation of $m_{T}$ is,
\begin{eqnarray}\nonumber
m_{T}&=&(m_{T})_{\Sigma}(\frac{r}{r_{\Sigma}})^3-r^3\int_{r}^{r_{\Sigma}}\frac{e^{(\nu+\lambda)/2}}{\tilde{r}}\\\label{a36}
&\times&\left[\frac{8\pi}{H}(P^{(eff)}_{\bot}-P^{(eff)}_{r})+\frac{1}{\tilde{r}^3}\int_{0}^{r}4\pi
\tilde{r}^3\left(\frac{\mu^{(eff)}}{H}\right)'d\tilde{r}\right]d\tilde{r}.
\end{eqnarray}
Using the information from Eq.\eqref{a29}, one can write
\begin{equation}\label{a37}
m_{T}=(m_{T})_{\Sigma}(\frac{r}{r_{\Sigma}})^3-r^3\int_{r}^{r_{\Sigma}}\frac{e^{(\nu+\lambda)/2}}
{\tilde{r}}[\frac{4\pi}{H}(P^{(eff)}_{\bot}-P^{(eff)}_{r})-E]d\tilde{r}.
\end{equation}
This relation could be helpful to understand the role of Weyl
scalar, modified correction terms, effective pressure anisotropy
and irregularity in the energy density of the static spherically
symmetric spacetime on the Tolman mass. Thus relates the phenomenon
of the occurrence of inhomogeneous energy density and local
anisotropy of pressure through Tolman mass in
$f(R,T,R_{\mu\nu}T^{\mu\nu})$ gravity.

\section{The Orthogonal Splitting of The Riemann Tensor}

The orthogonal splitting of Riemann curvature tensor was proposed by
Bel \cite{bel1961inductions} and Herrera \emph{et al.}
\cite{herrera2004spherically}. One can find three tensors obtained
from the orthogonal decomposition of the Riemann tensor, as
\begin{align}\label{a38}
Y_{\alpha\beta}&=R_{\alpha\gamma\beta\delta}u^\gamma u^\delta,
\\\label{a39}
Z_{\alpha\beta}&=*R_{\alpha\gamma\beta\delta}u^\gamma
u^\delta=\frac{1}{2}\eta_{\alpha\gamma\epsilon\mu}R^{\epsilon\mu}_{\beta\delta}u^\gamma
u^\delta,
\\\label{a40}
X_{\alpha\beta}&=*R^*_{\alpha\gamma\beta\delta}u^\gamma
u^\delta=\frac{1}{2}\eta_{\alpha\gamma}^{\epsilon\mu}R^*_{\epsilon\mu\beta\delta}u^\gamma
u^\delta,
\end{align}
where $\eta_{\alpha\gamma}^{\epsilon\mu}$ represents the Levi-Civita
symbol while the steric indicates the dual operation on the
subsequent tensor. One can rewrite the Riemann tensor in terms of
above mentioned tensors (see \cite{gomez2007dynamical}), therefore
following this we have calculated another form of Eq.\eqref{a23}
after using field equation as
\begin{equation}\label{a41}
R^{\alpha\gamma}_{\quad\beta\delta}=C^{\alpha\gamma}_{\quad\beta\delta}+16\pi
T^{(eff)[\alpha}_{[\beta} \delta_{\delta]}^{\gamma]}+8\pi
T^{(eff)}\left(\frac{1}{3}\delta^{\alpha}_{\quad[\beta}
\delta_{\delta]}^{\gamma}-\delta^{[\alpha}_{\quad[\beta}
\delta_{\delta]}^{\gamma]}\right),
\end{equation}
which further can be written through Eq.\eqref{a3} as
\begin{equation}\label{a42}
R^{\alpha\gamma}_{\quad\beta\delta}=R^{\alpha\gamma}_{(I)\beta\delta}+R^{\alpha\gamma}_{(II)\beta\delta}+R^{\alpha\gamma}_{(III)\beta\delta},
\end{equation}
where
\begin{eqnarray}\nonumber
  R^{\alpha\gamma}_{(I)\beta\delta}&=&\frac{16\pi}{H}(f_{T}+\frac{1}{2}Rf_{Q}+1) \left[\mu u^{[\alpha}u_{[\beta} \delta_{\delta]}^{\gamma]}- Ph^{[\alpha}_{\quad[\beta} \delta_{\delta]}^{\gamma]}+ \Pi^{[\alpha}_{\quad[\beta} \delta_{\delta]}^{\gamma]}\right]\\\nonumber
  &+&\frac{8\pi}{H}\left[(f_{T}+\frac{1}{2}Rf_{Q}+1)(\mu-3P)+4\left\{\frac{R}{2}\left(\frac{f}{R}-f_{R}\right)+\mu f_{T}\right.\right.\\\nonumber
  &-&\left.\frac{1}{2}\nabla_{\mu}\nabla_{\nu}(f_{Q}T^{\mu\nu})\right\}-\frac{1}{2}\Box\{f_{Q}(\mu-3P)\}-3\Box f_{R}-2f_{Q}R_{\mu\alpha}T^{\mu\alpha}\\\nonumber
  &+&\left.\nabla_{\mu}\nabla_{\alpha}(f_{Q}T^{\mu\alpha})+2g^{\alpha\xi}(f_{Q}R^{\mu\nu}+f_{T}g^{\mu\nu})\frac{\partial^2L_{m}}{\partial g^{\alpha\xi}\partial g^{\mu\nu}}\right]\\\label{a43}
  &\times&\left(\frac{1}{3}\delta^{\alpha}_{\quad[\beta} \delta_{\delta]}^{\gamma}-\delta^{[\alpha}_{\quad[\beta} \delta_{\delta]}^{\gamma]}\right),
\end{eqnarray}
\begin{eqnarray}\nonumber
  R^{\alpha\gamma}_{(II)\beta\delta}&=&\frac{4\pi}{H}\left[2\left\{\frac{R}{2}\left(\frac{f}{R}-f_{R}\right)+\mu f_{T}-\frac{1}{2}\nabla_{\mu}\nabla_{\nu}(f_{Q}T^{\mu\nu})\right\}\right.\\\nonumber
  &\times&\left(\delta^{\alpha}_{\beta} \delta_{\delta}^{\gamma}-\delta^{\alpha}_{\delta} \delta_{\beta}^{\gamma}\right)-\frac{1}{2}\Box\left\{f_{Q}\left(T^{\alpha}_{\beta} \delta_{\delta}^{\gamma}-T^{\alpha}_{\delta} \delta_{\beta}^{\gamma}-T^{\gamma}_{\beta} \delta_{\delta}^{\alpha}+T^{\gamma}_{\delta} \delta_{\beta}^{\alpha}\right)\right\}\\\nonumber
  &-&2\Box f_{R}\left(\delta^{\alpha}_{\beta} \delta_{\delta}^{\gamma}-\delta^{\alpha}_{\delta} \delta_{\beta}^{\gamma}\right)+\left( \delta_{\delta}^{\gamma}\nabla^{\alpha}\nabla_{\beta}- \delta_{\beta}^{\gamma}\nabla^{\alpha}\nabla_{\delta}- \delta_{\delta}^{\alpha}\nabla^{\gamma}\nabla_{\beta}\right.\\\nonumber
  &+&\left. \delta_{\beta}^{\alpha}\nabla^{\gamma}\nabla_{\delta}\right)f_{R}
  -f_{Q}\left(R^{\alpha}_{\mu}T^{\mu}_{\beta} \delta_{\delta}^{\gamma}-R^{\alpha}_{\mu}T^{\mu}_{\delta} \delta_{\beta}^{\gamma}-R^{\gamma}_{\mu}T^{\mu}_{\beta} \delta_{\delta}^{\alpha}+R^{\gamma}_{\mu}T^{\mu}_{\delta} \delta_{\beta}^{\alpha}\right)\\\nonumber
  &-&f_{Q}\left(R_{\mu\beta}T^{\mu\alpha} \delta_{\delta}^{\gamma}-R_{\mu\delta}T^{\mu\alpha} \delta_{\beta}^{\gamma}
  -R_{\mu\beta}T^{\mu\gamma} \delta_{\delta}^{\alpha}+R_{\mu\delta}T^{\mu\gamma} \delta_{\beta}^{\alpha}\right)\\\nonumber
  &+&\frac{1}{2}\nabla_{\mu}\nabla^{\alpha}\left\{f_{Q}\left(T^{\mu}_{\beta} \delta_{\delta}^{\gamma}-T^{\mu}_{\delta} \delta_{\beta}^{\gamma}\right)\right\}+\frac{1}{2}\nabla_{\mu}\nabla_{\beta}\left\{f_{Q}\left(T^{\alpha\mu} \delta_{\delta}^{\gamma}-T^{\gamma\mu} \delta_{\delta}^{\alpha}\right)\right\}\\\nonumber
  &+&\frac{1}{2}\nabla_{\mu}\nabla^{\gamma}\left\{f_{Q}\left(T^{\mu}_{\delta} \delta_{\beta}^{\alpha}-T^{\mu}_{\beta} \delta_{\delta}^{\alpha}\right)\right\}+\frac{1}{2}\nabla_{\mu}\nabla_{\delta}\left\{f_{Q}\left(T^{\gamma\mu} \delta_{\beta}^{\alpha}-T^{\alpha\mu} \delta_{\beta}^{\gamma}\right)\right\}\\\nonumber
  &+&2g^{\alpha\gamma}(f_{Q}R^{\mu\nu}+f_{T}g^{\mu\nu})\left\{\delta_{\delta}^{\gamma}\frac{\partial^2L_{m}}{\partial g^{\gamma\beta}\partial g^{\mu\nu}}-\delta_{\beta}^{\gamma}\frac{\partial^2L_{m}}{\partial g^{\gamma\delta}\partial g^{\mu\nu}}\right.\\\label{a44}
  &-&\delta_{\delta}^{\alpha}\frac{\partial^2L_{m}}{\partial g^{\alpha\beta}\partial g^{\mu\nu}}+\left.\left.\delta_{\beta}^{\alpha}\frac{\partial^2L_{m}}{\partial g^{\alpha\delta}\partial g^{\mu\nu}}\right\}\right],\\\label{a45}
  R^{\alpha\gamma}_{(III)\beta\delta}&=&4u^{[\alpha}u_{[\beta} E_{\delta]}^{\gamma]}-\epsilon^{\alpha\gamma}_{\mu}\epsilon_{\beta\delta\nu}E^{\mu\nu},
\end{eqnarray}
with
\begin{equation}\label{a46}
\epsilon_{\alpha\gamma\beta}=u^\mu \eta_{\mu\alpha\gamma\beta},\quad
\epsilon_{\alpha\gamma\beta}u^\beta=0,
\end{equation}
where we have used that the magnetic part of Conformal tensor is
identically zero in case of spherical symmetry.

In order to find three important tensors, i.e., $X_{\alpha\beta},
Y_{\alpha\beta}$ and $Z_{\alpha\beta}$, we use above mentioned
equation and get
\begin{eqnarray}\nonumber
Y_{\alpha\beta}&=&E_{\alpha\beta}+\frac{1}{H}\left\{\frac{4\pi}{3}(\mu+3P)h_{\alpha\beta}+4\pi
\Pi_{\alpha\beta}\right\}(f_{T}+\frac{1}{2}Rf_{Q}+1)\\\nonumber
&-&\frac{8\pi}{3H}\left\{\frac{R}{2}\left(\frac{f}{R}-f_{R}\right)+\mu
f_{T}-\frac{1}{2}\nabla_{\mu}\nabla_{\nu}(f_{Q}T^{\mu\nu})\right\}h_{\alpha\beta}\\\nonumber
&+&\frac{4\pi}{H}\left[-\frac{1}{2}\left\{\Box(f_{Q}T_{\alpha\beta})-u_{\beta}u^{\delta}\Box(f_{Q}T_{\alpha\delta})\right.\right.\\\nonumber
&-&\left.u_{\alpha}u_{\gamma}\Box(f_{Q}T^{\gamma}_{\beta})
+g_{\alpha\beta}u_{\gamma}u^{\delta}\Box(f_{Q}T^{\gamma}_{\delta})\right\}\\\nonumber
&+&\left(\nabla_{\alpha}\nabla_{\beta}f_{R}-u_{\beta}u^{\delta}\nabla_{\alpha}\nabla_{\delta}
f_{R}-u_{\alpha}u_{\gamma}\nabla^{\gamma}\nabla_{\beta}f_{R}
+g_{\alpha\beta}u_{\gamma}u^{\delta}\nabla^{\gamma}\nabla_{\delta}f_{R}\right)\\\nonumber
&+&f_{Q}\left\{R_{\alpha\mu}(Ph^{\mu}_{\beta}-\Pi^{\mu}_{\beta})-R^{\gamma}_{\mu}(\mu
u^{\mu}u_{\gamma}h_{\alpha\beta}+u_{\alpha}u_{\gamma}Ph^{\mu}_{\beta}-u_{\alpha}u_{\gamma}\Pi^{\mu}_{\beta})\right\}\\\nonumber
&+&f_{Q}\left\{R_{\mu\beta}(Ph^{\mu}_{\alpha}-\Pi^{\mu}_{\alpha})-R_{\mu\delta}(\mu
u^{\mu}u^{\delta}h_{\alpha\beta}+u_{\beta}u^{\delta}Ph^{\mu}_{\alpha}-u_{\beta}u^{\delta}\Pi^{\mu}_{\alpha})\right\}\\\nonumber
&+&\frac{1}{2}\{\nabla_{\mu}\nabla_{\alpha}(f_{Q}T^{\mu}_{\beta})+\nabla_{\mu}\nabla_{\beta}(f_{Q}T^{\mu}_{\alpha})
+g_{\alpha\beta}u_{\gamma}u^{\delta}\nabla_{\mu}\nabla^{\gamma}(f_{Q}T^{\mu}_{\delta})\\\nonumber
&+&g_{\alpha\beta}u_{\gamma}u^{\delta}\nabla_{\mu}\nabla_{\delta}(f_{Q}T^{\mu\gamma})-u_{\beta}
u^{\delta}\nabla_{\mu}\nabla_{\alpha}(f_{Q}T^{\mu}_{\delta})
-u_{\gamma}u_{\alpha}\nabla_{\mu}\nabla_{\beta}(f_{Q}T^{\gamma\mu})\\\nonumber
&-&u_{\gamma}u_{\alpha}\nabla_{\mu}\nabla^{\gamma}(f_{Q}T^{\mu}_{\beta})-u_{\beta}u^{\delta}
\nabla_{\mu}\nabla_{\delta}(f_{Q}T^{\mu}_{\alpha})\}\\\nonumber
&+&\left.2h^{\epsilon}_{\alpha}(f_{Q}R^{\mu\nu}+f_{T}g^{\mu\nu})\frac{\partial^2L_{m}}{\partial
g^{\epsilon\beta}\partial
g^{\mu\nu}}\right]+\frac{8\pi}{3H}\left[\frac{1}{2}\Box\{f_{Q}(\mu-3P)\}\right.\\\nonumber
&+&2f_{Q}R_{\mu\epsilon}(\mu
u^{\mu}u^{\epsilon}-Ph^{\mu\epsilon}+\Pi^{\mu\epsilon})
-\nabla_{\mu}\nabla_{\epsilon}(f_{Q}T^{\mu\epsilon})\\\label{a47}
&-&\left.2g^{\epsilon\xi}(f_{Q}R^{\mu\nu}+f_{T}g^{\mu\nu})\frac{\partial^2L_{m}}{\partial
g^{\epsilon\xi}\partial g^{\mu\nu}}\right]h_{\alpha\beta},
\end{eqnarray}
\begin{eqnarray}\nonumber
  Z_{\alpha\beta}&=&\frac{4\pi}{H}\left[\frac{1}{2}u^{\delta}\Box(f_{Q}T^{\epsilon}_{\delta})
  -u^{\delta}\nabla^{\epsilon}\nabla_{\delta}f_{R}+f_{Q}\mu R_{\delta}^{\epsilon}u^{\delta}
  -f_{Q}PR_{\delta}^{\epsilon}u^{\delta}+\frac{1}{3}f_{Q}\Pi R^{\epsilon}_{\delta}u^{\delta}\right.\\\label{a48}
  &-&\left.\frac{1}{2}u^{\delta}\nabla_{\mu}\nabla^{\epsilon}(f_{Q}T^{\mu}_{\delta})
  -\frac{1}{2}u^{\delta}\nabla_{\mu}\nabla_{\delta}(f_{Q}T^{\mu\epsilon})\right]\epsilon_{\epsilon\beta\alpha},
\end{eqnarray}
and
\begin{eqnarray}\nonumber
X_{\alpha\beta}&=&-E_{\alpha\beta}+\frac{1}{H}\left(\frac{8\pi}{3}\mu
h_{\alpha\beta}+4\pi
\Pi_{\alpha\beta}\right)(f_{T}+\frac{1}{2}Rf_{Q}+1)\\\nonumber
&+&\frac{4\pi}{H}\left[\left\{-\frac{1}{2}\Box(f_{Q}T^{\rho}_{\epsilon})+\nabla^{\rho}\nabla_{\epsilon}f_{R}
+\frac{1}{2}\nabla_{\mu}\nabla^{\rho}(f_{Q}T^{\mu}_{\epsilon})
\right.\right.\\\nonumber
&+&\left.\left.\frac{1}{2}\nabla_{\mu}\nabla_{\epsilon}(f_{Q}T^{\mu\rho})\right\}
\epsilon^{\epsilon\delta}_{\alpha}\epsilon_{\rho\delta\beta}+f_{Q}R^{\rho}_{\mu}\left(P-\frac{\Pi}{3}\right)
\epsilon^{\mu\delta}_{\alpha}\epsilon_{\rho\delta\beta}
\right.\\\nonumber
&+&\left.f_{Q}R_{\mu\epsilon}\left(P-\frac{\Pi}{3}\right)\epsilon^{\epsilon\delta}_{\alpha}\epsilon^{\mu}_{\delta\beta}\right]
+\frac{8\pi}{3H}\left[\left\{\frac{R}{2}\left(\frac{f}{R}-f_{R}\right)+\mu
f_{T}\right.\right.\\\nonumber
&-&\left.\frac{1}{2}\nabla_{\mu}\nabla_{\nu}(f_{Q}T^{\mu\nu})\right\}-\frac{1}{2}\Box
\{f_{Q}(\mu-3P)\}+2Rf_{Q}\left(P-\frac{\Pi}{3}\right)\\\label{a49}
&+&\left.\nabla_{\mu}\nabla_{\rho}(f_{Q}T^{\mu\rho})+2g^{\rho\epsilon}(f_{Q}R^{\mu\nu}+f_{T}g^{\mu\nu})\frac{\partial^2L_{m}}{\partial
g^{\rho\epsilon}\partial g^{\mu\nu}}\right]h_{\alpha\beta}.
\end{eqnarray}

Finally, we can obtain four structure scalars $X_{T}, X_{TF}, Y_{T}$
and $Y_{TF}$ from $X_{\alpha\beta}$ and $Y_{\alpha\beta}$, as
\begin{align}\label{a50}
X_{T}&=\frac{8\pi\mu}{H}(f_{T}+\frac{1}{2}Rf_{Q}+1)+\chi_{1}^{(D)},
\\\label{a51}
X_{TF}&=-E+\frac{4\pi\Pi}{H}(f_{T}+\frac{1}{2}Rf_{Q}+1),
\\\label{a53}
Y_{T}&=\frac{4\pi}{H}(\mu+3P_{r}-2\Pi)(f_{T}+\frac{1}{2}Rf_{Q}+1)+\chi_{2}^{(D)},
\\\label{a54}
Y_{TF}&=E+\frac{4\pi\Pi}{H}(f_{T}+\frac{1}{2}Rf_{Q}+1)+\chi_{3}^{(D)}.
\end{align}
On utilizing Eq.\eqref{a29} in Eq.\eqref{a51}, we obtain
\begin{equation}\label{a52}
X_{TF}=\frac{4\pi}{r^3}\int_{0}^{r}\tilde{r}^3\left(\frac{\mu^{(eff)}}{H}\right)'d\tilde{r}-\frac{4\pi\Pi^{(eff)}}{H}
+\frac{4\pi\Pi}{H}(f_{T}+\frac{1}{2}Rf_{Q}+1),
\end{equation}
where
$\chi_{3}^{(D)}=\frac{1}{s_{\alpha}s_{\beta}+\frac{1}{3}h_{\alpha\beta}}\chi_{\alpha\beta}^{(D)}$,
while the values of $\chi_{1}^{(D)}$, $\chi_{2}^{(D)}$ and
$\chi_{\alpha\beta}^{(D)}$ are given in Appendix B. Again utilizing
Eq.\eqref{a29}, one can write Eq.\eqref{a54} as
\begin{equation}\label{a55}
Y_{TF}=-\frac{4\pi}{r^3}\int_{0}^{r}\tilde{r}^3\left(\frac{\mu^{(eff)}}{H}\right)'d\tilde{r}+\frac{4\pi\Pi^{(eff)}}{H}
+\frac{4\pi\Pi}{H}(f_{T}+\frac{1}{2}Rf_{Q}+1) +\chi_{3}^{(D)}.
\end{equation}
The anisotropic pressure plus $f(R,T,Q)$ corrections in terms of
trace-free parts of structure scalars can be expressed, as
\begin{equation}\label{a56}
  X_{TF}+Y_{TF}=\frac{8\pi\Pi}{H}(f_{T}+\frac{1}{2}Rf_{Q}+1)+\chi_{3}^{(D)}.
\end{equation}

To demonstrate the realistic meaning of $Y_{TF}$, we utilize
Eq.\eqref{a55} in Eq.\eqref{a36} to yield
\begin{eqnarray}\nonumber
m_{T}&=&(m_{T})_{\Sigma}(\frac{r}{r_{\Sigma}})^3+r^3\int_{r}^{r_{\Sigma}}\frac{e^{(\nu+\lambda)/2}}{\tilde{r}}\left[Y_{TF}
+\frac{4\pi\Pi^{(eff)}}{H}\right.\\\label{a57}
&-&\left.\frac{4\pi\Pi}{H}(f_{T}+\frac{1}{2}Rf_{Q}+1)-\chi_{3}^{(D)}\right]
d\tilde{r}.
\end{eqnarray}
The matching the above equation with Eq.\eqref{a36}, we conclude
that $Y_{TF}$ provides the effects of the local pressure anisotropy
and inhomogeneous energy density on the Tolman mass under the effect
of extra curvature terms in $f(R,T,Q)$ theory. Moreover, the Tolman
mass can be expressed in an alternative way as
\begin{eqnarray}\nonumber
m_{T}&=&\int_{0}^{r}\tilde{r}^2e^{(\nu+\lambda)/2}\left[Y_{T}-\frac{4\pi}{H}(\mu+3P_{r}-2\Pi)(f_{T}
+\frac{1}{2}Rf_{Q}+1)\right.\\\label{a58}
&+&\left.\frac{4\pi}{H}(\mu^{(eff)}+3P_{r}^{(eff)}-2\Pi^{(eff)})-\chi_{2}^{(D)}\right]d\tilde{r}.
\end{eqnarray}
This equation has directly relate $Y_T$ to the effective matter
variables and Tolman mass function. It is well-known from the
working of Herrera \emph{et al.}
\cite{herrera2011meaning,Herrera2012} and Yousaf \emph{et al.}
\cite{PhysRevD.95.024024,bhatti2017dynamical,bhatti2017evolution}
that $Y_T$ has been involved in the evolutionary equation of the
expansion scalar which is widely known as Raychaudhuri equation.
Equation \eqref{a58} suggests that $m_T$ could be used to define the
Raychaudhuri equation even in $f(R,T,Q)$ gravity.

\section{Matter Distribution With Zero Complexity Factor}

It is well-known that there are many components which could be
reason to produce complexity in any static/non-static systems. The
causes of complexity in our structure are the inhomogeneous energy
density and pressure anisotropy contained in the structure scalar
$Y_{TF}$ in the modified gravity. The corresponding equations of
motion contain five unknowns $(\nu,~P_{r},~\lambda,~P_{\bot},~\mu)$
with $f(R,T,Q)$ theory. Therefore, in order to proceed our work, we
need two more conditions. One of them can be obtained from the
vanishing of complexity factor condition. After putting $Y_{TF}=0$
in Eq.\eqref{a55}, we get
\begin{equation}\label{a59}
\Pi=\frac{H}{2}\left[\frac{1}{r^3}\int_{0}^{r}\tilde{r}^3\left(\frac{\mu^{(eff)}}{H}\right)'d\tilde{r}
-\frac{\Pi^{(D)}}{H}-\frac{\Pi}{H}(f_{T}+\frac{1}{2}Rf_{Q})
-\frac{1}{4\pi}\chi_{3}^{(D)}\right].
\end{equation}
This equation can be noticed as a non-local equation of state as it
describes the radial pressure as a function of the energy density at
a specific point on the manifold (under the constraint
$f(R,T,Q)=R$). Now, we consider two subcases as follows:

\subsection{The Gokhroo and Mehra Ans\"{a}tz}

In order to continue the systematic analysis, we take the assumption
introduced by Gokhroo and Mehra, as
\begin{equation}\label{a60}
e^{-\lambda}=1-\alpha r^2+\frac{3K\alpha r^4}{5r_{\Sigma}^2},
\end{equation}
where $K\in(0,1)$, $\alpha=\frac{8\pi\mu_{o}}{3}$ and $\mu_{o}$ is a
constant. Utilizing Eq.\eqref{a60} in Eqs.\eqref{a11} and
\eqref{a18}, we get
\begin{equation}\label{a61}
\frac{\mu^{(eff)}}{H}=\mu_{o}\left(1-\frac{Kr^2}{r_{\Sigma}^2}\right),
\end{equation}
and
\begin{equation}\label{a62}
m(r)=\frac{4\pi\mu_{o}r^3}{3}\left(1-\frac{3Kr^2}{5r_{\Sigma}^2}\right).
\end{equation}
Further, from Eqs.\eqref{a12} and \eqref{a13}, we obtain
\begin{equation}\label{a63}
\frac{8\pi}{H}[P_{r}^{(eff)}-P_{\bot}^{(eff)}]=e^{-\lambda}\left[-\frac{\nu''}{2}
-\left(\frac{\nu'}{2}\right)^2+\frac{\nu'}{2r}+\frac{1}{r^2}+\frac{\lambda'}{2}
\left(\frac{\nu'}{2}+\frac{1}{r}\right)\right]-\frac{1}{r^2}.
\end{equation}
It could be useful to introduce new structural variables as
\begin{equation}\label{a64}
e^{\nu(r)}=e^{\int(2z(r)-2/r)dr},\quad e^{-\lambda(r)}=y(r),
\end{equation}
In this context, after substituting Eq.\eqref{a64} in
Eq.\eqref{a63}, we obtain
\begin{equation}\label{a66}
y'+y\left[\frac{2z'}{z}+2z-\frac{6}{r}+\frac{4}{r^2z}\right]=-\frac{2}{z}\left(\frac{1}{r^2}
+\frac{8\pi}{H}(P_{r}^{(eff)}-P_{\bot}^{(eff)})\right),
\end{equation}
which appears to be in form proposed by Ricatti. The integration of
this expression produces the line element in terms of $z$ and $\Pi$
in the background of $f(R,T,Q)$ theory as
\begin{eqnarray}\nonumber
ds^2&=&-e^{\int(2z(r)-2/r)dr}dt^2+\frac{z^2(r)e^{\int
\left(\frac{4}{r^2z(r)}+2z(r)\right)dr}}{r^6\left(-2\int\frac{z(r)\{1+8\pi\Pi^{(eff)}(r)r^2/H\}e^{\int
\left(\frac{4}{r^2z(r)}+2z(r)\right)dr}}{r^8}dr+C\right)}dr^2\\\label{a67}
&+&r^2(d\theta^2+sin^2\theta d\phi^2),
\end{eqnarray}
where $C$ is integration constant. Thus, one may write the effective
forms of physical variables as
\begin{align}\label{a68}
&\frac{4\pi P^{(eff)}_{r}}{H}=\frac{z(r-2m)+m/r-1}{r^2},
\\\label{a69}
&\frac{4\pi \mu^{(eff)}}{H}=\frac{m'}{r^2},
\\\label{a70}
&\frac{8\pi
P^{(eff)}_{\bot}}{H}=\left(1-\frac{2m}{r}\right)\left(z'+z^2-\frac{z}{r}
+\frac{1}{r^2}\right)+z\left(\frac{m}{r^2}-\frac{m'}{r}\right).
\end{align}
Such kind of solutions could be helpful to understand some hidden
and very interesting features of the spherical systems. Di Prisco
\emph{et al}. \cite{di2011expansion} calculated such results in the
background of GR. These results were modified by Sharif and Yousaf
for the cases of cylindrically \cite{sharif2012expansion} and
shearfree spherical \cite{sharif2012shearfree} structures. Yousaf
\cite{yousaf2017spherical} and Bhatti \cite{bhatti2016shear} further
extended these results for Einstein-$\Lambda$ gravity. To obtain
non-singular characteristics of the obtained solutions, the obtained
solutions \eqref{a68}-\eqref{a70} should satisfy
Eqs.\eqref{a20}-\eqref{27n2} at the boundary.

\subsection{The Polytropic Fluid With Zero Complexity Factor}

In this subsection, we study polytropic relativistic fluid in the
the presence of $f(R,T,Q)$ corrections. To deal with the system of
equations, we require vanishing complexity factor condition to
supplement with the polytropic equation of state.  Here, we consider
two cases of polytropes separately. The first one is
\begin{equation}\label{a71}
P^{(eff)}_{r}=K[\mu^{(eff)}]^{\gamma}=K[\mu^{(eff)}]^{(1+1/n)};
\quad Y_{TF}=0,
\end{equation}
with the polytropic constant $K$, polytropic exponent $\gamma$ and
polytropic index $n$.

To make the dimensionless form of TOV equation and the mass
function, we may define some new variables as
\begin{align}\label{a72}
\alpha&=P^{(eff)}_{rc}/\mu^{(eff)}_{c},\quad r=\xi/A,\quad
A^2=4\pi\mu^{(eff)}_{c}/\alpha(n+1),
\\\label{a73}
\psi^n&=\mu^{(eff)}/\mu^{(eff)}_{c},\quad
\nu(\xi)=m(r)A^3/(4\pi\mu^{(eff)}_{c}),
\end{align}
where subscript $c$ shows that the quantity is calculated at the
center. At the boundary $r=r_{\Sigma}(\xi=\xi_{\Sigma})$, we have
$\psi(\xi_{\Sigma})=0$. Putting these dimensionless variables in TOV
equation, we get
\begin{eqnarray}\nonumber
&&\xi^2\frac{d\psi}{d\xi}\left[\frac{1-2(n+1)\alpha\nu/\xi}{1+\alpha\psi}\right]+\frac{2\Pi^{(eff)}
\psi^{-n}\xi}{P^{(eff)}_{rc}(n+1)}\left[\frac{1-2(n+1)\alpha\nu/\xi}
{1+\alpha\psi}\right]\\\nonumber
&&+\nu+\alpha\xi^3\frac{\psi^{n+1}}{H}=\frac{\xi^2\psi^{-n}}{AP_{rc}^{(eff)}(n+1)}\left(\frac{1-2(n+1)
\alpha\nu/\xi}{1+\alpha\psi}\right)\\\label{a74}
&&\times\left[Ze^\lambda
H+\frac{A}{H}\frac{dH}{d\xi}K\psi^{n+1}(\mu_{c}^{(eff)})^{(1+1/n)}\right].
\end{eqnarray}
Differentiating $\nu(\xi)$ and substituting the value of $m(r)$, we
get
\begin{equation}\label{a75}
\frac{d\nu}{d\xi}=\frac{\xi^2\psi^n}{H}.
\end{equation}

In two ordinary differential equations \eqref{a74} and \eqref{a75},
there are three unknown functions $\nu,~\psi$ and $\Pi$. For the
unique solution of these equations, we need an extra condition. To
obtain this, we assume the vanishing complexity factor condition in
form of dimensionless variables. This gives
\begin{eqnarray}\nonumber
\frac{6\Pi}{n\mu^{(eff)}_{c}}&+&\frac{2\xi}{n\mu^{(eff)}_{c}}\frac{d\Pi}{d\xi}=\psi^{n-1}\xi\frac{d\psi}{d\xi}
+\frac{2\Pi\xi}{Hn\mu^{(eff)}_{c}}\frac{dH}{d\xi}\\\nonumber
&-&\frac{3}{n\mu^{(eff)}_{c}}\left[\Pi^{(D)}+\Pi(f_{T}+\frac{1}{2}Rf_{Q})+\frac{1}{4\pi
H}\chi_{3}^{(D)}\right]\\\nonumber
&+&\frac{H\xi}{n\mu^{(eff)}_{c}}\left[\mu^{(eff)}_{c}\psi^n\frac{d}{d\xi}\left(\frac{1}{H}\right)
-\frac{d}{d\xi}\left(\frac{\Pi^{(D)}}{H}\right)\right.\\\label{a76}
&-&\left.\frac{d}{d\xi}\left\{\frac{\Pi}{H}(f_{T}+\frac{1}{2}Rf_{Q})\right\}-\frac{1}{4\pi}\frac{d\chi_{3}^{(D)}}{d\xi}\right].
\end{eqnarray}
These differential equations have a unique solution for arbitrary
values of two parameters $n$ and $\alpha$. Any solution of these
equations provide the mass, density, radius and pressure of peculiar
stellar object for the specified values of $n$ and $\alpha$.

Now, we consider the second case of polytropic equation of state
$P^{(eff)}_{r}=K[\mu_{b}^{(eff)}]^{\gamma}=K[\mu_{b}^{(eff)}]^{(1+1/n)}$,
where $\mu_{b}$ is the baryonic mass density. Equations \eqref{a74}
and \eqref{a76} can be written in this case as
\begin{eqnarray}\nonumber
&&\xi^2\frac{d\psi_{b}}{d\xi}\left[\frac{1-2(n+1)\alpha\nu/\xi}{1+\alpha\psi_{b}}\right]+\frac{2\Pi^{(eff)}
\psi_{b}^{-n}\xi}{P^{(eff)}_{rc}(n+1)}\left[\frac{1-2(n+1)\alpha\nu/\xi}
{1+\alpha\psi_{b}}\right]\\\nonumber
&&+\nu+\alpha\xi^3\frac{\psi_{b}^{n+1}}{H}=\frac{\xi^2\psi_{b}^{-n}}{AP_{rc}^{(eff)}(n+1)}\left(\frac{1-2(n+1)
\alpha\nu/\xi}{1+\alpha\psi_{b}}\right)\\\label{a77}
&&\times\left[Ze^\lambda
H+\frac{A}{H}\frac{dH}{d\xi}K\psi_{b}^{n+1}(\mu_{c}^{(eff)})^{(1+1/n)}\right],
\end{eqnarray}
\begin{eqnarray}\nonumber
\frac{6\Pi}{n\mu^{(eff)}_{bc}}&+&\frac{2\xi}{n\mu^{(eff)}_{bc}}\frac{d\Pi}{d\xi}=\psi_{b}^{n-1}\xi
\frac{d\psi_{b}}{d\xi}\{1+K(n+1)(\mu_{bc}^{(eff)})^{1/n}\psi_{b}\}\\\nonumber
&+&\frac{2\Pi\xi}{Hn\mu^{(eff)}_{bc}}\frac{dH}{d\xi}-\frac{3}{n\mu^{(eff)}_{bc}}\left[\Pi^{(D)}+\Pi(f_{T}+\frac{1}{2}Rf_{Q})+\frac{1}{4\pi
H}\chi_{3}^{(D)}\right]\\\nonumber
&+&\frac{H\xi}{n\mu^{(eff)}_{bc}}\left[\mu^{(eff)}_{bc}\psi_{b}^n\{1+nK(\mu_{bc}^{(eff)})^{1/n}\psi_{b}\}\frac{d}{d\xi}\left(\frac{1}{H}\right)
-\frac{d}{d\xi}\left(\frac{\Pi^{(D)}}{H}\right)\right.\\\label{a78}
&-&\left.\frac{d}{d\xi}\left\{\frac{\Pi}{H}(f_{T}+\frac{1}{2}Rf_{Q})\right\}
-\frac{1}{4\pi}\frac{d\chi_{3}^{(D)}}{d\xi}\right],
\end{eqnarray}
with $\psi_{b}^n=\mu_{b}^{(eff)}/\mu_{bc}^{(eff)}$.

\section{Conclusions}

The aim of this work is to understand the effects of $f(R,T,Q)$ gravity on the structure of the self-gravitating spherical object. For this purpose, we have assumed static form of the spherical metric and then assumed that it is coupled with anisotropic matter configurations. The corresponding field as well as hydrostatic equilibrium equations are derived in the realm of $f(R,T,Q)$ theory. After using formalisms provided by Misner-Sharp and Tolman, particular relations of $m$ and $m_T$ are derived, respectively. Five set of $f(R,T,Q)$ scalar variables are derived from the orthogonal decomposition of the Riemann tensor. We then studied the impact of these variables in the emergence and maintenance of homogeneous distribution of matter content over the static relativistic spheres. Herrera \cite{herrera2018new} presented the concept of the complexity factor for static anisotropic self-gravitating spherically symmetric structure. The fundamental supposition is that the system with homogeneous energy density and anisotropic pressure is less complex. Then, the one of the derived scalar factor $Y_{TF}$ is explored in our case in order to determine the complexity factor of the system. Now, we are going to point out some important results as follows.\\

(i) In background of $f(R,T,Q)$ theory, the factor $Y_{TF}$ involves inhomogeneous energy density and locally anisotropic pressure under the influence of modified corrections.\\

(ii) The quantity $Y_{TF}$ measures the Tolman mass in terms of inhomogeneous energy density and anisotropic pressure with the contribution of extra curvature terms of modified gravity.\\

(iii) This scalar could contain the dissipative fluxes with inhomogeneous energy density and pressure anisotropy in non-static dissipative matter distribution with extra curvature terms of modified gravity.\\

(iv) A new definition of complexity for spherically symmetric static self-gravitating fluids has been introduced \cite{herrera2018new}, which is sharply different from the definition given in \cite{de2012entropy}. Indeed, the new concept of complexity \cite{herrera2018new}, stems from the basic assumption that one of the less complex systems corresponds to a homogeneous (in the energy density) fluid distribution with isotropic pressure. So a zero value of the complexity factor is assigned for such a distribution. Then, as an obvious candidate to measure the degree of complexity, emerges a quantity which appears in the orthogonal splitting of the Riemann tensor and that was denoted by$Y_{TF}$ and called the complexity factor.\\

After establishing the modified field equations and matter function,
we obtain the complexity factor $Y_{TF}$ given in Eq.\eqref{a55}
from the four structure scalars. This scalar has been arrived from
the orthogonal splitting of the Riemann tensor which encompasses
pressure anisotropy and density inhomogeneity. Furthermore, we
propose two applications of physical systems through the vanishing
complexity factor condition \eqref{a59} by taking $Y_{TF}=0$. The
first one is Gokhroo and Mehra ans\"{a}tz, in which we have observed
the effects of extra curvature terms in stellar objects. In second
example, we have considered the polytropic equation of state and
introduced new variables to write TOV equation, vanishing complexity
factor condition and mass function in dimensionless form. To explain
the system, the above differential equations yield the solution for
some physical constraints with zero complexity factor under the
effect of modified corrections. All of our results reduce to GR
\cite{herrera2018new} under the constraint $f(R,T,Q)=R$.

\vspace{0.25cm}

\section*{Appendix A}

The effective matter variables appearing in
Eqs.\eqref{a11}-\eqref{a13} are
\begin{eqnarray}\nonumber
\mu^{(eff)}&=&\mu\left[1+2f_{T}+f_{Q}\left(\frac{1}{2}R-\frac{3\nu'^2}{8e^{\lambda}}-\frac{3\nu'}{2re^{\lambda}}+\frac{5\lambda'\nu'}{8e^{\lambda}}
-\frac{3\nu''}{4e^{\lambda}}\right)\right.\\\nonumber
&+&\left.f'_{Q}\left(\frac{1}{re^{\lambda}}-\frac{\lambda'}{4e^{\lambda}}\right)+\frac{f''_{Q}}{2e^{\lambda}}\right]
+\mu'\left[f_{Q}\left(\frac{1}{re^{\lambda}}-\frac{\lambda'}{4e^{\lambda}}\right)+\frac{f'_{Q}}{e^{\lambda}}\right]
+\frac{\mu''f_{Q}}{2e^{\lambda}}\\\nonumber
&+&P_{r}\left[f_{Q}\left(\frac{\nu'^2}{8e^{\lambda}}-\frac{1}{r^2e^{\lambda}}-\frac{\lambda'\nu'}{8e^{\lambda}}+\frac{\lambda'}{2re^{\lambda}}
+\frac{\nu''}{4e^{\lambda}}\right)+f'_{Q}\left(\frac{\lambda'}{4e^{\lambda}}-\frac{2}{re^{\lambda}}\right)\right.\\\nonumber
&-&\left.\frac{f''_{Q}}{2e^{\lambda}}\right]+P'_{r}\left[f_{Q}\left(\frac{\lambda'}{4e^{\lambda}}-\frac{2}{re^{\lambda}}\right)
-\frac{f'_{Q}}{e^{\lambda}}\right]-\frac{P''_{r}f_{Q}}{2e^{\lambda}}\\\nonumber
&+&P_{\bot}\left[f_{Q}\left(\frac{\nu'}{2re^{\lambda}}+\frac{1}{r^2e^{\lambda}}-\frac{\lambda'}{2re^{\lambda}}\right)
+\frac{f'_{Q}}{re^{\lambda}}\right]+\frac{P'_{\bot}f_{Q}}{re^{\lambda}}+\frac{R}{2}\left(\frac{f}{R}-f_{R}\right)\\\label{a79}
&+&f'_{R}\left(\frac{2}{re^{\lambda}}-\frac{\lambda'}{2e^{\lambda}}\right)+\frac{f''_{R}}{e^{\lambda}},
\end{eqnarray}
\begin{eqnarray}\nonumber
P_{r}^{(eff)}&=&\mu\left[-f_{T}+f_{Q}\left(\frac{\nu'^2}{8e^{\lambda}}+\frac{\nu'}{2re^{\lambda}}
-\frac{\lambda'\nu'}{8e^{\lambda}}+\frac{\nu''}{4e^{\lambda}}\right)
-\frac{f'_{Q}\nu'}{4e^{\lambda}}\right]-\frac{\mu'\nu'f_{Q}}{4}\\\nonumber
&+&P_{r}\left[1+f_{T}+f_{Q}\left(\frac{1}{2}R-\frac{3\nu'^2}{8e^{\lambda}}+\frac{\nu'}{re^{\lambda}}
+\frac{1}{r^2e^{\lambda}}+\frac{3\lambda'\nu'}{8e^{\lambda}}
+\frac{3\lambda'}{2re^{\lambda}}\right.\right.\\\nonumber
&-&\left.\frac{3\nu''}{4e^{\lambda}}\right)+f'_{Q}\left(\frac{1}{re^{\lambda}}-\frac{\nu'}{4e^{\lambda}}
+\left.\frac{\nu'}{2e^{\lambda}}\right)\right]+P'_{r}\left[f_{Q}\left(\frac{1}{re^{\lambda}}
+\frac{\nu'}{4e^{\lambda}}\right)\right]\\\nonumber
&+&P_{\bot}\left[f_{Q}\left(-\frac{\nu'}{2re^{\lambda}}-\frac{1}{r^2e^{\lambda}}+\frac{\lambda'}{2re^{\lambda}}
\right)+\frac{f'_{Q}}{re^{\lambda}}\right]+\frac{P'_{\bot}f_{Q}}{r}\\\label{a80}
&-&\frac{R}{2}\left(\frac{f}{R}-f_{R}\right)-f'_{R}\left(\frac{\nu'}{2e^{\lambda}}+\frac{2}{re^{\lambda}}\right),
\end{eqnarray}
\begin{eqnarray}\nonumber
P_{\bot}^{(eff)}&=&\mu\left[-f_{T}+f_{Q}\left(\frac{\nu'^2}{8e^{\lambda}}+\frac{\nu'}{2re^{\lambda}}
-\frac{\lambda'\nu'}{8e^{\lambda}}+\frac{\nu''}{4e^{\lambda}}\right)
+\frac{f'_{Q}\nu'}{4e^{\lambda}}\right]+\frac{\mu'\nu'f_{Q}}{4e^{\lambda}}\\\nonumber
&+&P_{r}\left[f_{Q}\left(\frac{\nu'^2}{8e^{\lambda}}+\frac{\nu'}{2re^{\lambda}}-\frac{\lambda'\nu'}{8e^{\lambda}}+\frac{\nu''}{4e^{\lambda}}\right)
+f'_{Q}\left(\frac{\nu'}{2e^{\lambda}}+\frac{1}{re^{\lambda}}-\frac{\lambda'}{4e^{\lambda}}\right)\right.\\\nonumber
&+&\left.\frac{f''_{Q}}{2e^{\lambda}}\right]+P'_{r}\left[f_{Q}\left(\frac{\nu'}{2e^{\lambda}}+\frac{1}{re^{\lambda}}
-\frac{\lambda'}{4e^{\lambda}}\right)+\frac{f'_{Q}}{e^{\lambda}}\right]
+\frac{P''_{r}f_{Q}}{2e^{\lambda}}\\\nonumber
&+&P_{\bot}\left[1+f_{T}+f_{Q}\left(\frac{1}{2}R-\frac{2}{r^2e^{\lambda}}+\frac{\lambda'}{re^{\lambda}}
-\frac{\nu'}{re^{\lambda}}+\frac{2}{r^2}\right)\right.\\\nonumber
&+&\left.f'_{Q}\left(\frac{\nu'}{4e^{\lambda}}-\frac{\lambda'}{4e^{\lambda}}
\right)+\frac{f''_{Q}}{2e^{\lambda}}\right]+P'_{\bot}\left[f_{Q}\left(\frac{\nu'}{4e^{\lambda}}
-\frac{\lambda'}{4e^{\lambda}}+\frac{2}{re^{\lambda}}\right)+\frac{f'_{Q}}{e^{\lambda}}\right]\\\label{a81}
&+&\frac{P''_{\bot}f_{Q}}{2e^{\lambda}}-\frac{R}{2}\left(\frac{f}{R}-f_{R}\right)
+f'_{R}\left(\frac{\lambda'}{2e^{\lambda}}-\frac{1}{re^{\lambda}}-\frac{\nu'}{2e^{\lambda}}\right)-\frac{f''_{R}}{e^{\lambda}}.
\end{eqnarray}

The quantity $Z$ arises due to non-conserved nature of $f(R,T,Q)$
theory in Eq.\eqref{a14} is
\begin{eqnarray}
  \nonumber Z &=& \frac{2}{\left(2+Rf_{Q}+2f_{T}\right)}\left[f'_{Q}e^{-\lambda}P_{r}\left(\frac{\nu'}{r}-\frac{e^\lambda}{r^2}+\frac{1}{r^2}\right)
  +f_{Q}e^{-\lambda}P_{r}\left(\frac{\nu''}{r}-\frac{\lambda'}{r^2}-\frac{\nu'\lambda'}{r}-\frac{\nu'}{r^2}\right.\right.\\\nonumber
  &+&\left.\frac{2e^\lambda}{r^3}-\frac{2}{r^3}\right)+\frac{f_{Q}e^{-\lambda}}{2}P_{r}'\left(\frac{\nu'\lambda'}{4}
  -\frac{\nu'^2}{4}-\frac{\nu''}{4}+\frac{\lambda'}{r}-\frac{f_{T}}{2}\right)-\mu f'_{T}-\mu'\left\{\frac{f_{Q}e^{-\lambda}}{8}\right.\\\nonumber
  &\times&\left.\left(-\nu'\lambda'+\nu'^2+2\nu''+\frac{4\nu'}{r}\right)
  +\frac{3f_{T}}{2}\right\}-P_{\bot}'\left\{\frac{f_{Q}e^{-\lambda}}{r}\left(\frac{\lambda'}{2}-\frac{\nu'}{2}\frac{e^\lambda}{r}
  -\frac{1}{r}\right)-f_{T}\right\}\\\label{a82}
  &-&\left.\left(\frac{1}{r^2}-\frac{e^{-\lambda}}{r^2}-\frac{\nu'e^{-\lambda}}{r}\right)\left(\mu'f_{Q}+\mu f'_{Q}\right)+P_{r}f'_{T}\right].
\end{eqnarray}

The quantity $D_0$ appearing in Eq.\eqref{a20} can be given as
follows
\begin{eqnarray}\nonumber
D_{0}&=&\mu\left[-\tilde{f}_{T}+\tilde{f}_{Q}\left(\frac{\nu'^2}{8e^{\lambda}}+\frac{\nu'}{2re^{\lambda}}
-\frac{\lambda'\nu'}{8e^{\lambda}}+\frac{\nu''}{4e^{\lambda}}\right)\right]-\frac{\mu'\nu'\tilde{f}_{Q}}{4}\\\nonumber
&+&P_{r}\left[\tilde{f}_{T}+\tilde{f}_{Q}\left(\frac{1}{2}R-\frac{3\nu'^2}{8e^{\lambda}}+\frac{\nu'}{re^{\lambda}}
+\frac{1}{r^2e^{\lambda}}+\frac{3\lambda'\nu'}{8e^{\lambda}}
+\frac{3\lambda'}{2re^{\lambda}}-\frac{3\nu''}{4e^{\lambda}}\right)\right]\\\nonumber
&+&P'_{r}\left[\tilde{f}_{Q}\left(\frac{1}{re^{\lambda}}
+\frac{\nu'}{4e^{\lambda}}\right)\right]+P_{\bot}\left[\tilde{f}_{Q}\left(-\frac{\nu'}{2re^{\lambda}}-\frac{1}{r^2e^{\lambda}}
+\frac{\lambda'}{2re^{\lambda}}\right)\right]\\\label{a83}
&+&\frac{P'_{\bot}\tilde{f}_{Q}}{r}-\frac{R}{2}\left(\frac{\tilde{f}}{R}-\tilde{f}_{R}\right).
\end{eqnarray}

\section*{Appendix B}

The extra curvature terms appearing in the expressions of structure
scalars \eqref{a50}-\eqref{a54} are found as follows
\begin{eqnarray}\nonumber
  \chi_{1}^{(D)}&=&\frac{4\pi}{H}\left[\left\{h^{\epsilon}_{\rho}\Box(f_{Q}T^{\rho}_{\epsilon})
  -2h^{\epsilon}_{\rho}\nabla^{\rho}\nabla_{\epsilon}f_{R}
-h^{\epsilon}_{\rho}\nabla_{\mu}\nabla^{\rho}(f_{Q}T^{\mu}_{\epsilon})
\right.\right.\\\nonumber
&-&\left.\left.h^{\epsilon}_{\rho}\nabla_{\mu}\nabla_{\epsilon}(f_{Q}T^{\mu\rho})\right\}
-2f_{Q}R^{\rho}_{\mu}\left(P-\frac{\Pi}{3}\right)h^{\mu}_{\rho}
-2f_{Q}R_{\mu\epsilon}\left(P-\frac{\Pi}{3}\right)h^{\epsilon\mu}\right]
\\\nonumber
&+&\frac{8\pi}{H}\left[\left\{\frac{R}{2}\left(\frac{f}{R}-f_{R}\right)+\mu
f_{T}-\frac{1}{2}\nabla_{\mu}\nabla_{\nu}(f_{Q}T^{\mu\nu})\right\}\right.\\\nonumber
&-&\frac{1}{2}\Box\{f_{Q}(\mu-3P)\}+2Rf_{Q}\left(P-\frac{\Pi}{3}\right)+\nabla_{\mu}\nabla_{\rho}(f_{Q}T^{\mu\rho})\\\label{a84}
&+&\left.2g^{\rho\epsilon}(f_{Q}R^{\mu\nu}+f_{T}g^{\mu\nu})\frac{\partial^2L_{m}}{\partial
g^{\rho\epsilon}\partial g^{\mu\nu}}\right],
\end{eqnarray}
\begin{eqnarray}\nonumber
\chi_{2}^{(D)}&=&-\frac{8\pi}{H}\left\{\frac{R}{2}\left(\frac{f}{R}-f_{R}\right)+\mu
f_{T}-\frac{1}{2}\nabla_{\mu}\nabla_{\nu}(f_{Q}T^{\mu\nu})\right\}+\frac{4\pi}{H}\left[-\frac{1}{2}\right.\\\nonumber
&\times&\left.\left\{\Box(f_{Q}T)-u^{\alpha}u^{\delta}\Box(f_{Q}T_{\alpha\delta})-u^{\beta}u_{\gamma}\Box(f_{Q}T^{\gamma}_{\beta})
+4u_{\gamma}u^{\delta}\Box(f_{Q}T^{\gamma}_{\delta})\right\}\right.\\\nonumber
&+&\left(\Box
f_{R}-u^{\alpha}u^{\delta}\nabla_{\alpha}\nabla_{\delta}f_{R}-u^{\beta}u_{\gamma}\nabla^{\gamma}\nabla_{\beta}f_{R}
+4u_{\gamma}u^{\delta}\nabla^{\gamma}\nabla_{\delta}f_{R}\right)\\\nonumber
&+&f_{Q}\left\{R^{\beta}_{\mu}(Ph^{\mu}_{\beta}-\Pi^{\mu}_{\beta})-3R^{\gamma}_{\mu}\mu
u^{\mu}u_{\gamma}\right\}
+f_{Q}\left\{R^{\alpha}_{\mu}(Ph^{\mu}_{\alpha}-\Pi^{\mu}_{\alpha})\right.\\\nonumber
&-&\left.3R_{\mu\delta}\mu
u^{\mu}u^{\delta}\right\}+\frac{1}{2}\{\nabla_{\mu}\nabla_{\alpha}(f_{Q}T^{\mu\alpha})+\nabla_{\mu}\nabla_{\beta}(f_{Q}T^{\mu\beta})
\\\nonumber
&+&4u_{\gamma}u^{\delta}\nabla_{\mu}\nabla^{\gamma}(f_{Q}T^{\mu}_{\delta})+4u_{\gamma}
u^{\delta}\nabla_{\mu}\nabla_{\delta}(f_{Q}T^{\mu\gamma})-u^{\alpha}u^{\delta}\nabla_{\mu}\nabla_{\alpha}(f_{Q}T^{\mu}_{\delta})
\\\nonumber
&-&u_{\gamma}u^{\beta}\nabla_{\mu}\nabla_{\beta}(f_{Q}T^{\gamma\mu})-u_{\gamma}
u^{\beta}\nabla_{\mu}\nabla^{\gamma}(f_{Q}T^{\mu}_{\beta})-u^{\alpha}u^{\delta}\nabla_{\mu}\nabla_{\delta}(f_{Q}T^{\mu}_{\alpha})\}\\\nonumber
&+&\left.2h^{\epsilon\beta}(f_{Q}R^{\mu\nu}+f_{T}g^{\mu\nu})\frac{\partial^2L_{m}}{\partial
g^{\epsilon\beta}\partial g^{\mu\nu}}\right]\\\nonumber
&+&\frac{8\pi}{H}\left[\frac{1}{2}\Box\{f_{Q}(\mu-3P)\}+2f_{Q}R_{\mu\epsilon}T^{\mu\epsilon}
-\nabla_{\mu}\nabla_{\epsilon}(f_{Q}T^{\mu\epsilon})\right.\\\label{a85}
&-&\left.2g^{\epsilon\xi}(f_{Q}R^{\mu\nu}+f_{T}g^{\mu\nu})\frac{\partial^2L_{m}}{\partial
g^{\epsilon\xi}\partial g^{\mu\nu}}\right],
\end{eqnarray}
\begin{eqnarray}\nonumber
\chi_{\alpha\beta}^{(D)}&=&-\frac{2\pi}{H}\left[h^{\lambda}_{\alpha}h^{\pi}_{\beta}\Box(f_{Q}T_{\lambda\pi})-\Box(f_{Q}T_{\alpha\beta})
-u_{\alpha}u_{\beta}u_{\gamma}u^{\delta}\Box(f_{Q}T^{\gamma}_{\delta})\right]\\\nonumber
&+&\frac{4\pi}{H}[(h^{\lambda}_{\alpha}h^{\pi}_{\beta}\nabla_{\pi}\nabla_{\lambda}f_{R}
-\nabla_{\alpha}\nabla_{\beta}f_{R}-u_{\alpha}u_{\beta}u_{\gamma}u^{\delta}\nabla^{\gamma}\nabla_{\delta}f_{R})
\\\nonumber
&+&f_{Q}(h^{\lambda}_{\alpha}h^{\mu}_{\beta}R_{\lambda\mu}P-h^{\mu}_{\beta}R_{\alpha\mu}P
-h^{\lambda}_{\alpha}R_{\lambda\mu}\Pi^{\mu}_{\beta}+R_{\alpha\mu}\Pi^{\mu}_{\beta})\\\nonumber
&+&f_{Q}(h^{\mu}_{\alpha}h^{\pi}_{\beta}R_{\mu\pi}P-h^{\mu}_{\alpha}R_{\mu\beta}P-h^{\pi}_{\beta}R_{\mu\pi}\Pi^{\mu}_{\alpha}
+R_{\mu\beta}\Pi^{\mu}_{\alpha})\\\nonumber
&+&\frac{1}{2}\{h^{\lambda}_{\alpha}h^{\pi}_{\beta}\nabla_{\mu}\nabla_{\lambda}(f_{Q}T^{\mu}_{\pi})
+h^{\lambda}_{\alpha}h^{\pi}_{\beta}\nabla_{\mu}\nabla_{\pi}(f_{Q}T^{\mu}_{\lambda})-\nabla_{\mu}\nabla_{\alpha}(f_{Q}T^{\mu}_{\beta})
\\\nonumber
&-&\nabla_{\mu}\nabla_{\beta}(f_{Q}T^{\mu}_{\alpha})-u_{\alpha}u_{\beta}u_{\gamma}u^{\delta}\nabla_{\mu}\nabla^{\gamma}(f_{Q}T^{\mu}_{\delta})
-u_{\alpha}u_{\beta}u_{\gamma}u^{\delta}\nabla_{\mu}\nabla_{\delta}(f_{Q}T^{\mu\gamma})\}\\\label{a86}
&+&2(f_{Q}R^{\mu\nu}+f_{T}R^{\mu\nu})h^{\epsilon}_{\alpha}\{h^{\pi}_{\beta}\frac{\partial^2L_{m}}{\partial
g^{\epsilon\pi}\partial g^{\mu\nu}} -\frac{\partial^2L_{m}}{\partial
g^{\epsilon\beta}\partial g^{\mu\nu}}\}].
\end{eqnarray}

\vspace{0.5cm}

{\bf Acknowledgments}

\vspace{0.25cm} The work of Z.Y., M.Z.B. and T.N. on static
solutions in the modified gravity was supported by National Research
Project for Universities (NRPU), Higher Education Commission,
Pakistan under the research project No.
8754/Punjab/NRPU/R\&D/HEC/2017. The work of M.Yu.K.\ on the
orthogonal splitting of the Riemann tensor in modified gravity was
supported by grant of the Russian Science Foundation (Project
No-18-12-00213).

\end{document}